\documentclass[12pt]{iopart}

\usepackage{epsfig}

\begin{document}

\title[Symmetry Aspects in Nonrelativistic Multi-Scalar Field Models  and 
Application]
{Symmetry Aspects in Nonrelativistic Multi-Scalar Field Models  and 
Application to a Coupled Two-Species Dilute Bose Gas\footnote{Based on a 
talk given by 
R. O. Ramos at the QFEXT05 workshop, Barcelona, Spain, September 5-9, 2005}}

\author{Rudnei O Ramos$^1$ and Marcus B Pinto$^2$}

\address{$^1$ Departamento de F\'{\i}sica
Te\'orica, Universidade do Estado do Rio de Janeiro, 20550-013 Rio
de Janeiro, RJ, Brazil} 

\address{$^2$ Departamento de
F\'{\i}sica, Universidade Federal de Santa Catarina, 88040-900
Florian\'{o}polis, SC, Brazil}

\eads{\mailto{rudnei@uerj.br}, \mailto{marcus@fsc.ufsc.br}}

\begin{abstract}
We discuss unusual aspects of symmetry that can happen due to entropic
effects in the context of multi-scalar field theories at finite
temperature. We present their consequences, in special, for the case of
nonrelativistic models of hard core spheres. We show that for
nonrelativistic models phenomena like inverse symmetry breaking and
symmetry non-restoration cannot take place, but a reentrant
phase at high temperatures
is shown to be possible for some region of parameters.
We then develop a model of interest
in studies of Bose-Einstein condensation in dilute atomic gases and
discuss about its phase transition patterns. In this application
to a Bose-Einstein condensation model, however, no reentrant phases 
are found.
\end{abstract}

\centerline{Published: J. Phys. A: Math. Gen. {\bf 39}, 6687 (2006)}

\section{Introduction}

One of the most interesting aspects concerning the studies of
multi-field models at finite temperature is the possibility of emergence
of a much richer phase diagram than one would usually find in one field
type of models. The possibility that unusual symmetry patterns could
emerge in those models, for some specific region of parameters, has
attracted considerable attention in the literature (see, for instance,
\cite{MR,onon} and references therein and \cite{qfext1}
for a short review).

{}For most of the standard physical systems we know in nature, we have 
a good sense of how symmetries seem to change as the temperature is
changed. Typically, the larger is the temperature the larger is the
symmetry exhibited by the system and vice-versa. Examples of this
behaviour are expected to happen in the particle physics models, 
like in the electroweak phase transition and possibly also in
Grand-Unified models. The same is expected in the much lower energy
systems, like those of condensed matter. {}For example, the phase
transition in ferromagnets, superconductors, Bose-Einstein condensation
of atomic gases, etc, just to name a few systems. In all these examples
we always go from an ordered (less symmetric) phase below some critical
temperature of phase transition to a disordered (more symmetrical) phase
above the critical temperature, or the opposite, if the temperature is
decreased from a high temperature (or symmetry restored) state
\cite{goldenfeld}.

However, the above symmetry aspects seem not to be the rule. In fact we
are also becoming increasingly aware that entropic effects in
multi-field models may show other patterns of symmetry breaking and
restoration that are less usual. {}For example, many condensed matter
systems, like spin glasses, compounds known as the manganites, liquid
crystals and many others, commonly show phenomena like reentrant phases
of lower symmetries at higher temperatures and, therefore, they can exhibit
unusual phase diagrams that we would otherwise not expect. Many of these
systems have recently been reviewed in \cite{inverse}. As concerned to
quantum field theory models, the possibility of other phase transition
patterns was also shown to be possible in the context of multi-scalar
field theories at finite temperatures \cite{MR}. These models show the
possibility of a symmetry that is not broken at low temperatures, 
getting broken at high temperatures (what is called an inverse symmetry
breaking). Other case that seems possible is a symmetry that is broken
at lower temperatures, not getting restored at all as we go to higher
temperatures (what is called a symmetry non-restoration). 
The problem of how a symmetry broken or restored phase may
emerge as a reentrant phase in the system, like it is seen in
many low energy condensed matter systems, was recently analyzed in
\cite{onon} in the context of a coupled nonrelativistic model of two
scalar fields.

The plan of this paper is as follows. In the next section we briefly
review the results obtained in \cite{onon} for the case of a
nonrelativistic model of two scalar fields with overall symmetry
$U(1)\times U(1)$ and show that it admits reentrant phases for some
region of parameters. One possible physical realization of this kind of
model is for example in the description of a coupled two species dilute
atomic gas system. This is a kind of system in atomic physics that has
been of great interest recently concerning studies (both theoretical and
experimental) of Bose-Einstein condensation. In Sec. 3 we offer a
quantum field theory description for this problem. We then analyze the 
possibility of emergence of reentrant phases in the quantum field
formulation for these systems. This would be a novel
symmetry behaviour that could be of great interest, given its possible
implementation in the laboratory. Our results allow us to conclude, at
the level of our approximations, on the non appearance of such reentrant
phases in these kind of coupled dilute atomic gases systems.

\section{Reentrant Phases in Nonrelativistic Multi-Scalar Field Models}

We start our discussion by considering the following
nonrelativistic Lagrangian density model, of two (complex) scalar fields
$\Phi$ and $\Psi$, with global symmetry $U(1) \times U(1)$,

\begin{eqnarray}
{\cal L}(\Phi^*,\Phi,\Psi^*,\Psi) &=&
\Phi^* \left(i \partial_t + \frac{1}{2m_\Phi}\nabla^2
\right) \Phi - \mu_\Phi \Phi^* \Phi 
- \frac{g_\Phi}{3!} (\Phi^* \Phi)^2
\nonumber \\
&+& \Psi^* \left(i \partial_t + \frac{1}{2m_\Psi}\nabla^2
\right) \Psi - \mu_\Psi \Psi^* \Psi 
- \frac{g_\Psi}{3!} (\Psi^* \Psi)^2
\nonumber \\
&-& g (\Phi^* \Phi) (\Psi^* \Psi)\;.
\label{NRL}
\end{eqnarray}

\noindent
The model (\ref{NRL}) can be thought as coming from the nonrelativistic
limit of a corresponding relativistic counterpart, as shown explicitly
in \cite{onon}. In (\ref{NRL}) the interaction parameters $g_\Phi$ and
$g_\Psi$ describe two-body self-interaction terms, as commonly
considered for dilute and cold (low energy) systems of particles
\cite{bec}, in which case only binary type interactions, {\it i.e.},
hard core type of interactions of the form shown in (\ref{NRL}), are
relevant. $g$ is the cross-coupling between the two fields, which we
consider here as a quadratic type of interaction. The mass parameters
$m_\Phi$ and $m_\Psi$ are the masses for the fields (or particles). The
one-body type of interactions, of magnitude $\mu_\Phi$ and $\mu_\Psi$,
can either represent the effect of external potentials (for example a
magnetic field) on the system, internal energy terms (like the internal
molecular energy relative to free atoms in which case the fields in the
Lagrangian would be related to molecular dimers), an explicit gap of
energy in the system (like in superconductors), or just chemical
potentials added to the action in the grand-canonical formulation to
enforce finite density (or fixed number of particles) for both $\Phi$
and $\Psi$. The latter will be the case for our application of
(\ref{NRL}) to the coupled atomic gas problem in Sec. 3. Here we will
just consider the $\mu_i$ parameters as constant one-body parameters
added to our model such that the possible symmetry breaking patterns,
depending on the sign of $\mu_i$ ($i=\Phi,\Psi$), can easily be
determined from the potential term in (\ref{NRL}). This is just what is
done in spontaneous symmetry breaking studies performed on the
relativistic analogous models. Therefore, for $\mu_i>0$, we have an
initially symmetry restored phase in both $\Phi$ and $\Psi$ directions,
while $\mu_i<0$ corresponds (at zero temperature) to symmetry broken
phases for both $\Phi$ and $\Psi$.

We require the model (\ref{NRL}) to be overall bounded from below, which
then gives the constraint condition on the two-body interaction terms,
$g_{\Psi}>0$, $g_{\Phi}>0$ and $g_{\Psi}g_{\Phi}>9g^2$. This is the same
condition imposed on the analogous relativistic problem \cite{MR,onon}.
This boundness constraint will be observed in all our results below.
Non-trivial phase transitions can emerge for negative values of the
cross-coupling $g$, which is allowed by the above boundness constraint.
This was shown in diverse instances to be the case in the relativistic
analogous models \cite{MR,qfext1}. 

In the analysis below, we will also restrict, for simplicity, to an
initial symmetry restored phase (at zero temperature) for both fields,
$\mu_i >0$ and leave the symmetry broken case for the Bose-Einstein
condensation problem studied in Sec. 3. The phase structure of the model
is then determined by the sign of the temperature dependent one-body
terms, $\mu_i(T)= \mu_i + \Sigma_i$, where $\Sigma_i$ is the field
temperature dependent self-energy. We look for an intermediate
(reentrant) phase at some interval of temperature and parameters in
which the symmetry in one of the field directions is broken. This
analysis can in principle be carried out within perturbation theory, as
described in a companion paper \cite{qfext1}, which indeed shows the
possibility of appearance of reentrant phases for a system of hard-core
particles described by (\ref{NRL}). However, the results in \cite{onon,qfext1}
also shows that as the (perturbative) temperature
corrections are considered for the two-body terms, $g_\Phi(T),
g_\Psi(T)$, these effective couplings run to negative values above
some temperature $T_{\rm neg}$. This then violates the initial condition of 
boundness
for the potential when considering the model in equilibrium in a thermal
bath with temperature $T>T_{\rm neg}$. 
At the leading order
perturbative calculation and assuming $\mu_\Phi = \mu_\Psi=\mu$ and
$m_\Phi = m_\Psi=m$, $T_{\rm neg} \simeq {\rm min} \left( 12 \pi
\sqrt{\mu/(2 m^3)} g_\Phi/(5 g_\Phi^2 + 9g^2),\; 12 \pi \sqrt{\mu/(2
m^3)} g_\Psi/(5 g_\Psi^2 + 9g^2) \right)$. This is reminiscent of the
breakdown of perturbation theory in quantum field theory at finite
temperature, which is well known in relativistic models (see e.g. \cite{GR} and
references in there). Nonperturbative methods are then called for a
proper interpretation of the results and to confirm that the appearance
of reentrant phases in our model is not just an artifact of perturbation
theory.

The problem of the self-couplings running to negative values at high
temperatures can be solved, e.g., by resumming all leading order bubble
corrections to the couplings. This is naturally done in the context of
the renormalization group, by solving the flow equations for all
couplings and parameters of the model. A simpler and equivalent approach
was also shown in \cite{onon}, where this resummation is also
accomplished by solving a set of self-consistent homogeneous linear
equations for all effective couplings, $g_\Phi(T), g_\Psi(T)$ and
$g(T)$, and the result of these equations feeded back in the equations for
the effective one-body terms, $\mu_i(T)$. We refer the interested reader
to \cite{onon} for the details and we give here only the main results of
this approach.

We consider, for illustrative purposes, the parameters (at
$T=0$): $g_{\Phi}=2\times 10^{-15} {\rm eV}^{-2}, g=- 10^{-16}{\rm
eV}^{-2}$, $m_\Phi\simeq m_\Psi=1\,{\rm GeV}$ and $\mu_\Phi=\mu_\Psi =
1\,{\rm neV}$. These values of couplings and masses could for instance
be representative of some dilute Bose gas atom or molecule (see next
section). The temperature and the tree-level value of $g_\Psi$ are then
changed and we look for regions of symmetry broken phase in the $\Psi$
field direction (for the values of parameters considered, it easy to
show that the symmetry remains always restored in the $\Phi$ direction
\cite{qfext1}). {}For all values of parameters and temperature
considered we check the boundness condition extended for the effective
couplings (temperature dependent), as obtained by the flow equations
described previously. That is, $g_{\Psi}(T) g_{\Phi}(T) > 9 g^2(T)$, or
that the ratio $R^{\rm NR} (T) = g_{\Psi}(T) g_{\Phi}(T)/[9 g^2(T)] >1$.
The resulting phase diagram, as a function of $g_\Psi$ at $T=0$ and 
the (log of the) temperature, is shown in {}Fig. 1. 

\begin{figure}[htb]
\vspace{0.5cm}
\centerline{\epsfig{figure=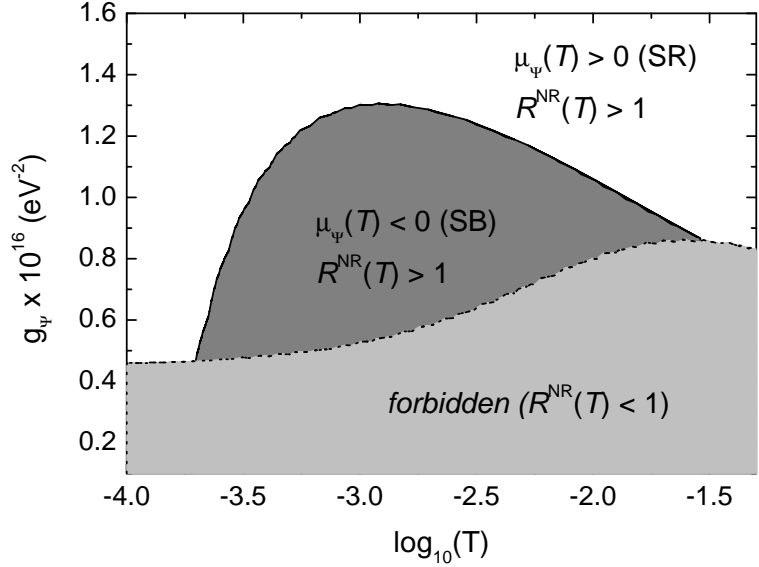,angle=0,width=10cm}}
\caption[]{\label{phasediagram} The phase diagram of the system in terms
of $g_\Psi$ and temperature, for the parameters considered in the text.
The dark gray region denotes a reentrant phase with symmetry breaking (SB)
in the $\Psi$ direction, $\mu_\Psi(T) <0$. 
The region below it, in light gray, is the unstable region,
$R^{\rm NR}(T) <1$ and above it is a symmetry restored (SR) phase,
$\mu_\Psi(T) >1$. Temperature is given in units of eV.}
\end{figure}

{}Fig. 1 shows clearly the possibility
of reentrant phases in the system, through an inverse symmetry breaking, in
$\Psi$ direction. {}For instance, for  
$g_\Psi(T=0) = 10^{-16}{\rm
eV}^{-2}$, we find a reentrant symmetry broken phase starting at the
temperature $T_{c,\Psi}^{\rm (ISB)} \simeq 3.4 \times 10^{-4} {\rm eV}$ (or
$\sim 4$ K) and ending at $T_{c,\Psi}^{\rm (SR)} \simeq 1.4 \times 10^{-2}
{\rm eV}$ (or $\sim 161$ K), through symmetry restoration. In this
region of temperature and parameters, $R^{\rm NR}(T) >1$ and the
(effective) potential is still bounded from below. In the $\Phi$
direction there is no symmetry breaking or reentrant phases at any
temperature for the parameters considered.

\section{Application to a Coupled Two-Species Dilute and Homogeneous
Atomic Bose Gas Model}

Let us now consider the case of model (\ref{NRL}) as describing two
coupled Bose gases of fixed densities $\rho_\Phi$ and $\rho_\Psi$,
respectively. The model could then be describing a system composed by a
mixture of coupled atomic gases, like the ones recently produced \cite
{myatt}, with same chemical element in two different hyperfine states, or even
two different mono-atomic Bose gases in the homogeneous case
\cite{coupled}. Here $\mu_\Phi$ and $\mu_\Psi$ are then explicitly
chemical potentials added in the grand-canonical formalism to ensure the
fixed densities for each Bose atom gas. Here we start describing the
system in the broken phase in both $\Phi$ and $\Psi$ directions.
Therefore, $\mu_\Phi \to - \mu_\Phi$ and $\mu_\Psi \to - \mu_\Psi$ in
(\ref{NRL}) and these chemical potentials are taken as positive
quantities and with their values determined by the usual thermodynamic 
relation, in terms of the pressure $P(T,\mu_\Phi,\mu_\Psi)$,

\begin{eqnarray}
\rho_\Psi = \frac{\partial P(T,\mu_\Phi,\mu_\Psi)}{\partial \mu_\Psi}
\;,\;\;\;\;
\rho_\Phi = \frac{\partial P(T,\mu_\Phi,\mu_\Psi)}{\partial \mu_\Phi}
\;,
\label{densities}
\end{eqnarray}

\noindent
where the pressure is defined as the negative of the effective potential
computed at its minima (which is the thermodynamic free energy of the
system),

\begin{equation}
P\equiv P(T,\mu_\Phi,\mu_\Psi) = - V_{\rm eff} (T,\phi_0,\psi_0) 
\Bigr|_{\phi_0 = \phi_m, \psi_0=\psi_m}\;,
\label{pressure}
\end{equation}

\noindent
where $\phi_m$ and $\psi_m$ are the values of $\phi_0$ and $\psi_0$ that
extremizes (corresponding to a minimum of) the effective potential,

\begin{eqnarray}
\frac{\partial V_{\rm eff} (T,\phi_0,\psi_0)}{\partial \phi_0}
\Bigr|_{\phi_0 = \phi_m, \psi_0=\psi_m} =0\;,\;\;\;\;
\frac{\partial V_{\rm eff} (T,\phi_0,\psi_0)}{\partial \psi_0}
\Bigr|_{\phi_0 = \phi_m, \psi_0=\psi_m} =0\;.
\label{minimum}
\end{eqnarray}

\noindent
The effective potential follows from (\ref{NRL}) by expanding the fields
around the vacuum expectation values, $\langle \Phi \rangle =
\phi_0/\sqrt{2}$ and $\langle \Psi \rangle = \psi_0/\sqrt{2}$, and it is
evaluated in the one-loop approximation in a standard computation of
quantum field theory at finite temperature (for the one-field case, see
for instance \cite{norway}). At the tree-level, $\phi_m$ and $\psi_m$
are given in terms of the minima of the potential in (\ref{NRL}), 

\begin{equation}
\phi_m^2= 6 \frac{g_\Psi \mu_\Phi - 3g \mu_\Psi}{g_\Psi g_\Phi - 9g^2}  \;,
\;\;\;
\psi_m^2= 6 \frac{g_\Phi \mu_\Psi - 3g \mu_\Phi}{g_\Psi g_\Phi - 9g^2}\;.
\label{phimpsim}
\end{equation}
At finite temperature, the equations for  $\phi_m$ and $\psi_m$ are 
given by analogous expressions to (\ref{phimpsim}), but in terms of the
effective chemical potentials instead, 
$\bar{\mu}_\Phi$ and $\bar{\mu}_\Psi$, that are defined by the solution of
the self-consistent equations,
$\bar{\mu}_\Phi = \mu_\Phi - \Sigma_{\phi,\phi}$ and 
$\bar{\mu}_\Psi = \mu_\Psi - \Sigma_{\psi,\psi}$, given in terms of the 
$\Phi$ and $\Psi$ field self-energies $\Sigma$.

The explicit expression for the pressure at finite temperature, obtained
from the effective potential as described above and that follows from
some length but straightforward calculation, is given by \cite{CMR}

\begin{eqnarray}
P(T,\mu_\Phi, \mu_\Psi) &=&  \frac{3}{ 2 \left( g_\Phi g_\Psi - 9 g^2 \right) }
\left[ g_\Psi \left( \mu_\Phi^2 -
\Sigma^2_{\phi,\phi} \right) + g_\Phi \left( \mu_\Psi^2 -
\Sigma^2_{\psi,\psi} \right) \right. \nonumber \\
&+& \left. 6 g \left( \Sigma_{\phi,\phi}
\Sigma_{\psi,\psi} - \mu_\Phi \mu_\Psi \right) \right]
- \int \frac {d^3 {\bf q}}{(2 \pi)^3} \left( \bar{A}_+ + \bar{A}_- \right)
\Bigr|_{ {\lower 1.5pt\hbox{\tiny $\psi_0=\psi_m$} 
\above 0pt \raise 1.5pt\hbox{\tiny $\phi_0=\phi_m$}}}
\nonumber \\
&-& \frac{1}{\beta} \int \frac {d^3 {\bf q}}{(2 \pi)^3} \left[
\ln \left( 1- e^{- \beta \bar{A}_+ } \right) 
+ \ln \left( 1- e^{- \beta \bar{A}_- } \right) \right]
\Bigr|_{ {\lower 1.5pt\hbox{\tiny $\psi_0=\psi_m$} 
\above 0pt \raise 1.5pt\hbox{\tiny $\phi_0=\phi_m$}}}\;,
\label{Pphipsi}
\end{eqnarray}

\noindent
where

\begin{equation}
\bar{A}_{\pm}^2 = \frac{\bar{H}_\Psi \bar{G}_\Psi + \bar{H}_\Phi \bar{G}_\Phi}{2} \mp
\frac{1}{2} \left[ \left( \bar{H}_\Psi \bar{G}_\Psi - \bar{H}_\Phi \bar{G}_\Phi \right)^2 +
4 g^2 \psi_m^2 \phi_m^2 \bar{G}_\Psi \bar{G}_\Phi \right]^{1/2} ,
\label{A+-}
\end{equation}
and

\begin{eqnarray}
\bar{H}_\Psi \!\!=\!\!  \frac{{\bf q}^2}{2 m_\Psi} + \frac{g_\Psi}{3} \psi_m^2\;,\;\;\;
\bar{G}_\Psi \!\!=\!\!  \frac{{\bf q}^2}{2 m_\Psi} \;, \;\;\;
\bar{H}_\Phi \!\!=\!\!  \frac{{\bf q}^2}{2 m_\Phi} + \frac{g_\Phi}{3} \phi_m^2\;, \;\;\;
\bar{G}_\Phi \!\!=\!\!  \frac{{\bf q}^2}{2 m_\Phi} \;.
\label{HG}
\end{eqnarray}

\noindent
$\bar{H}_i$ and $\bar{G}_i$ denote the Higgs and Goldstone modes,
respectively, for each field in the broken phase.

{}From (\ref{densities}) and after some algebra to eliminate the
dependence of these expressions on the chemical potentials, we find the
expressions relating the total densities $\rho_\Psi$ and $\rho_\Phi$
with (the condensate densities) $\psi_m$ and $\phi_m$, as given by
\cite{CMR}

\begin{equation}
\rho_\Psi = \frac{\psi_m^2}{2} - \frac{1}{2} \int \frac {d^3 {\bf q}}{(2 \pi)^3}
\left[ \frac{\partial\bar{A}_+} {\partial \mu_\Psi} \left( 1 + 2 n_{\bar{A}_+} \right)
+ \frac{\partial\bar{A}_-} {\partial \mu_\Psi} \left( 1 + 2 n_{\bar{A}_-} \right)
\right] \;,
\label{rhopsid2}
\end{equation}

\noindent
and 

\begin{equation}
\rho_\Phi = \frac{\phi_m^2}{2} - \frac{1}{2} \int \frac {d^3 {\bf q}}{(2 \pi)^3}
\left[ \frac{\partial\bar{A}_+} {\partial \mu_\Phi} \left( 1 + 2 n_{\bar{A}_+} \right)
+ \frac{\partial\bar{A}_-} {\partial \mu_\Phi} \left( 1 + 2 n_{\bar{A}_-} \right)
\right] \;,
\label{rhophid2}
\end{equation}

\noindent
where $n_{\bar{A}_{\pm}} = 1/[\exp(\beta A_{\pm}) -1]$ and 
the partial derivatives of $\bar{A}_{\pm}$ with respect to 
$\mu_\Phi$ and $\mu_\Psi$ are defined by

\begin{eqnarray}
&& \frac{\partial \bar{A}_+}{\partial \mu_\Phi} =
-\frac{\left( \bar{H}_\Phi + \bar{G}_\Phi \right)}{4 \bar{A}_+ } \left[
1+ \frac{ 
\left( \bar{H}_\Psi \bar{G}_\Psi - \bar{H}_\Phi \bar{G}_\Phi \right)
- 2 g^2 \phi_m^2 \psi_m^2 \bar{G}_\Psi }{
\sqrt{ \left( \bar{H}_\Psi \bar{G}_\Psi - \bar{H}_\Phi \bar{G}_\Phi \right)^2
+ 4 g^2 \phi_m^2 \psi_m^2 \bar{G}_\Phi \bar{G}_\Psi} } \right], 
\nonumber \\
&& \frac{\partial \bar{A}_+}{\partial \mu_\Psi} =
-\frac{ \left( \bar{H}_\Psi + \bar{G}_\Psi \right) }{4 \bar{A}_+ } \left[1 - 
\frac{
\left( \bar{H}_\Psi \bar{G}_\Psi - \bar{H}_\Phi \bar{G}_\Phi \right)
+ 2 g^2 \phi_m^2 \psi_m^2 \bar{G}_\Phi }{
\sqrt{ \left( \bar{H}_\Psi \bar{G}_\Psi - \bar{H}_\Phi \bar{G}_\Phi \right)^2
+ 4 g^2 \phi_m^2 \psi_m^2 \bar{G}_\Phi \bar{G}_\Psi} } \right] , 
\nonumber \\
&& \frac{\partial \bar{A}_-}{\partial \mu_\Phi} =
-\frac{ \left( \bar{H}_\Phi + \bar{G}_\Phi \right)}{4 \bar{A}_- } \left[ 1 -
 \frac{ \left( \bar{H}_\Psi \bar{G}_\Psi - \bar{H}_\Phi \bar{G}_\Phi \right)
- 2 g^2 \phi_m^2 \psi_m^2 \bar{G}_\Psi }{
\sqrt{ \left( \bar{H}_\Psi \bar{G}_\Psi - \bar{H}_\Phi \bar{G}_\Phi \right)^2
+ 4 g^2 \phi_m^2 \psi_m^2 \bar{G}_\Phi \bar{G}_\Psi} } \right], 
\nonumber \\
&& \frac{\partial \bar{A}_-}{\partial \mu_\Psi} =
-\frac{\left( \bar{H}_\Psi + \bar{G}_\Psi \right)}{4 \bar{A}_- } \left[ 1 +
\frac{
\left( \bar{H}_\Psi \bar{G}_\Psi - \bar{H}_\Phi \bar{G}_\Phi \right)
+ 2 g^2 \phi_m^2 \psi_m^2 \bar{G}_\Phi }{
\sqrt{ \left( \bar{H}_\Psi \bar{G}_\Psi - \bar{H}_\Phi \bar{G}_\Phi \right)^2
+ 4 g^2 \phi_m^2 \psi_m^2 \bar{G}_\Phi \bar{G}_\Psi} } \right]. 
\label{partialAB}
\end{eqnarray}

The coupled equations (\ref{rhopsid2}) and (\ref{rhophid2}) give
completely the phase diagram for the condensates $\psi_m$ and $\phi_m$
as a function of the temperature and the densities. A qualitative
analysis of the phase structure is also possible to be deduced already
at this level from equations (\ref{rhopsid2}), (\ref{rhophid2}) and
(\ref{partialAB}). Note that for $g=0$, from (\ref{A+-}) we obtain that
$\bar{A}_{+}^2(g=0) = \bar{H}_\Phi \bar{G}_\Phi$, $\bar{A}_{-}^2(g=0) =
\bar{H}_\Psi \bar{G}_\Psi$, and we obtain the Bogoliubov spectrum
\cite{bec} for each field in the uncoupled case. Also, (\ref{rhopsid2})
and (\ref{rhophid2}) decouples and we obtain as a result, for example
for $\rho_\Phi$,

\begin{eqnarray}
\rho_\Phi \!\!=\!\! \frac{\phi_m^2}{2} \!+\! \frac{1}{2} \!
\int \frac {d^3 {\bf q}}{(2 \pi)^3}  
\frac{\frac{{\bf q}^2}{2 m_\Phi} + \frac{g_\Phi \phi_m^2}{6}}{ \sqrt{ \frac{{\bf q}^2}{2 m_\Phi}
\left(\frac{{\bf q}^2}{2 m_\Phi} +\frac{g_\Phi \phi_m^2}{3}\right)} }
\! \!\left[ \! 1 \! +\! \frac{2}{e^{\beta \sqrt{ \frac{{\bf q}^2}{2 m_\Phi}
\left(\frac{{\bf q}^2}{2 m_\Phi} +
\frac{g_\Phi \phi_m^2}{3}\right)}} \!-1 } \right],
\label{rhophi4}
\end{eqnarray}

\noindent
with analogous equation for $\rho_\Psi$. Taking (\ref{rhophi4}) at the
critical point, $T=T_{c,\Phi}$, we have that $\phi_m(T=T_{c,\Phi}) =0$,
since the condensate density at $T_c$ vanishes and (\ref{rhophi4}) gives
$\rho_\Phi = \int d^3 {\bf q}/(2 \pi)^3 [\exp\left( {\bf q}^2/(2 m_\Phi
T_{c,\Phi})\right) -1 ]^{-1}$, or inverting it, $T_{c,\Phi} = 2
\pi/m_\Phi \left[\rho_\Phi/ \zeta(3/2) \right]^{2/3}$, where $\zeta(3/2)
\simeq 2.612$. This is the standard result for the critical temperature
of an homogeneous ideal Bose gas. This result emerges because of the
level of approximation we are considering. It is only modified by
corrections due to the self-interactions through nonperturbative methods
and it requires at least second order corrections in the self-energy
(see, for instance, \cite{shift} and references therein).

Note also that, at the level of approximation we are considering, from
the equations (\ref{A+-}), (\ref{rhopsid2}), (\ref{rhophid2}) and
(\ref{partialAB}), if any of the fields go above the transition point
(either $\phi_m=0$ or $\psi_m=0$) the two equations (\ref{rhopsid2}) and
(\ref{rhophid2}) also decouple, becoming independent of each other,
since the cross-coupling term in (\ref{A+-}) and (\ref{partialAB})
always appears multiplying both $\phi_m$ and $\psi_m$. As a result, no
reentrant phase at high temperatures seems to be possible here. A
computation performed in the restored phase case (similar to the one
done in Sec. 2) also seems to confirm this result. This comes about as a
consequence of the strong temperature dependence introduced by the
chemical potentials through the relation (\ref{densities}) in both the
broken and symmetric cases. In the broken (BEC) phase, it is also seen
that the system exhibits a small dependence on the cross-coupling term (at
one-loop order it is even insensitive to the sign of $g$). A throughout
analysis of the phase structure coming from the coupled set of equations
(\ref{rhopsid2}) and (\ref{rhophid2}), including higher order terms,
will be presented elsewhere \cite{CMR}.

\ack

The authors were partially supported by CNPq-Brazil. ROR was also 
partially supported by FAPERJ. We also would like
the organizers of the QFEXT05 workshop in Barcelona, Spain, for the
enjoyable conference atmosphere where this work has been presented.

\end{document}